\documentclass[conference]{IEEEtran}
\IEEEoverridecommandlockouts
\usepackage{cite}
\usepackage{amsmath,amssymb,amsfonts}
\usepackage{algorithmic}
\usepackage{graphicx}
\usepackage{textcomp}
\usepackage{xcolor}
\usepackage[hidelinks]{hyperref}
\usepackage{nomencl}
\usepackage{etoolbox}
\def\BibTeX{{\rm B\kern-.05em{\sc i\kern-.025em b}\kern-.08em
    T\kern-.1667em\lower.7ex\hbox{E}\kern-.125emX}}

\makenomenclature

\begin{document}

\nomenclature{$\Phi$}{Cost function}
\nomenclature{$y$}{Measurement vector}
\nomenclature{$Y$}{Output constraint set}
\nomenclature{$u$}{Set point vector}
\nomenclature{$U$}{Input constraint set}
\nomenclature{$h$}{Steady-state input to output map}
\nomenclature{$p$}{Number of set points}
\nomenclature{$n$}{Number of measurements}
\nomenclature{$q$}{Number of input constraints}
\nomenclature{$l$}{Number of output constraints}
\nomenclature{$\Delta h(u)$}{Sensitivity matrix}
\nomenclature{$\alpha$}{Controller gain}
\nomenclature{$\hat{\sigma}_{\alpha}$}{Optimization result, optimal step}
\nomenclature{$w$}{Optimization variable}
\nomenclature{$F$}{Set of controllable devices}
\nomenclature{$B$}{Set of observed lines}
\nomenclature{$N$}{Set of observed buses}
\nomenclature{$P_{set}$}{Active power set point}
\nomenclature{$P_{planned}$}{Original active power set point}
\nomenclature{$P_{meas}$}{Measured active power}
\nomenclature{$Q_{set}$}{Reactive power set point}
\nomenclature{$P_{max/min}$}{Maximum/Minimum active power}
\nomenclature{$Q_{max/min}$}{Maximum/Minimum reactive power}
\nomenclature{$V_{meas}$}{Measured bus voltage}
\nomenclature{$I_{meas}$}{Measured line current}
\nomenclature{$V_{max/min}$}{Maximum/Minimum bus voltage}
\nomenclature{$I_{max/min}$}{Maximum/Minimum line current}

\title{Tuning a Cascaded Online Feedback Optimization Controller for Provision of Distributed Flexibility \\
}

\author{\IEEEauthorblockN{Irina Zettl \IEEEauthorrefmark{1}, Florian Klein-Helmkamp \IEEEauthorrefmark{1}, Florian Schmidtke \IEEEauthorrefmark{1}\IEEEauthorrefmark{2}, Lukas Ortmann \IEEEauthorrefmark{3}, Andreas Ulbig\IEEEauthorrefmark{1}\IEEEauthorrefmark{2}}
	\IEEEauthorblockA{\IEEEauthorrefmark{1}\textit{IAEW at RWTH Aachen University, Aachen, Germany}}
	\IEEEauthorblockA{\IEEEauthorrefmark{2}\textit{Fraunhofer Center Digital Energy, Fraunhofer FIT, Aachen}}
	\IEEEauthorblockA{\IEEEauthorrefmark{3}\textit{OST - Eastern Switzerland University of Applied Sciences, Rapperswil, Switzerland} \\ E-mail: i.zettl@iaew.rwth-aachen.de}}

\maketitle

\begin{abstract}
Coordinating a high number of flexibility providing units (e.g. to provide ancillary services for the transmission system) across various grid layers requires new control concepts. A flexibility request at a point of common coupling can be met by utilizing a cascaded control structure based on online feedback optimization. In this paper the influence of the parameterization of the individual controllers on the performance of the hierarchical flexibility provision is studied on a three-level test system. The results show a high interdependency between the choice of control parameters of one controller and the behavior of other controllers as well as a significant impact on the accuracy and speed of flexibility provision. With a careful tuning, a cascaded structure based on online feedback optimization can achieve efficient vertical coordination of flexibility providing units.
 \\
\end{abstract}


\vspace{-3mm}
\printnomenclature[1.75cm]

\section{Introduction}
\label{sec:intro}
The replacement of large conventional power plants with small-scale, volatile generation units leads to higher fluctuations in the utilization of grid infrastructure and assets as well as bidirectional power flows\cite{iea2023}. 
A higher utilization of existing grid infrastructure presents a cost-effective alternative to grid reinforcement and expansion~\cite{innosys2030}. This can be realized through certain ancillary services, e.g. curative congestion management.  
Small-scale distributed energy resources (DER) connected to the distribution grid can collectively provide flexibility for such ancillary services. Their spatial distribution as well as their connection to various voltage levels motivate the research into new coordination concepts \cite{hoffrichter2019, graupner2022, frueh2023, kolster2020}.  

FPUs can be connected to various voltage levels within the distribution or transmission system. To utilize the provided flexibility, a coordination across multiple grid layers is necessary. This can also require communication and information exchange between different system operators. Conventionally, solutions based on optimal power flow (OPF) are used for this type of problem \cite{contreras2021}. These approaches rely on accurate grid models, which are not always available. One approach that allows vertical coordination of multiple FPUs with limited need for information exchange is a cascaded control algorithm based on online feedback optimization (OFO). OFO integrates an optimization algorithm in closed loop with the physical grid, tracking the optimal solution of the dispatch problem. The feedback structure of the controller allows the consideration of real-time measurements and the actual grid status for the optimization making the controller robust to disturbances and inaccurate grid models \cite{klein-helmkamp2024}.

\begin{figure}[tb]
	\centerline{\includegraphics{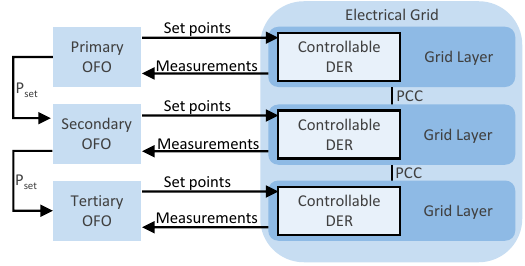}}
	\caption{Cascaded structure of individual OFO controllers}
	\label{fig:ofo_simple}
\end{figure}

Incorporated into a cascaded structure one or more OFO controllers can be used for each grid layer to provide the requested active power flow at the point of common coupling (PCC) with the superordinate grid layer (see Fig.~\ref{fig:ofo_simple}). This allows a bottom-up aggregation of flexibility.
A careful tuning of the controllers is necessary, so that the advantages of the algorithm can be used efficiently for vertical FPU coordination in a cascaded structure. 

\subsection{Related Work}
Distribution grids offer valuable flexibility but require careful coordination of DER and between various system operators \cite{kolster2022}. Coordination algorithms based on feed-forward optimization have been proposed in \cite{frueh2023} and \cite{kalantar2021} among others. Their drawbacks including the need for an accurate system model and lacking robustness against model mismatch are addressed by employing feedback control schemes with online optimization. 
Previous research has been conducted, exploring the theoretical foundations of feedback optimization concerning its stability and robustness. While \cite{molzahn2017, he2020, haberle2020, colombino2020, hauswirth2024, colombino2019, krishnamoorty2022} provide insights on feedback optimization in general, in \cite{picallo2022, hauswirth2017, ortmann2024a, picallo2020, ortmann2023} the utilization in the context of power system operation and flexibility provision from distributed resources has been studied.
    
In addition to theoretical considerations, experimental validations of the OFO control scheme have been conducted: In \cite{ortmann2020} voltage control with OFO was tested on a real distribution grid and compared to a dispatch based on optimal power flow. The ability of OFO to coordinate multiple DER for flexibility provision has been experimentally validated in \cite{klein-helmkamp2023} using a laboratory setup. A cascaded structure allows the flexibility provision and coordination across various grid layers and system operators. This has been shown in \cite{klein-helmkamp2024} but without focus on the parameterization of the individual controllers. The tuning of  OFO controllers in the context of curative flexibility provision was tested in \cite{ortmann2024b}, but limited to a single controller. Considering the whole cascade when choosing parameters can be relevant for efficient operation.

\subsection{Main Contribution}
In this paper, a cascaded control structure based on online feedback optimization for the coordination of decentralized flexibility providing units (FPUs) is investigated. For coordination across various grid layers and between different system operators, a cascaded control structure is implemented. Each level of the cascade is assigned to a specific grid layer and consists of one or more individual OFO controllers that observe their own area of the grid. The structure allows a top-down disaggregation of a flexibility request and a bottom-up aggregation of the requested flexibility. 
To achieve a fast and reliable change in active power flow at the PCC between different grid areas an attentive tuning of the controller cascade is necessary. The convergence behavior of the controllers can be influenced by changing the gain of the controllers. In this paper we investigate how the parameterization of the individual controllers in the cascade influences the performance of the flexibility provision to the overlying grid layer. For this we formulate the following research questions: 
\begin{enumerate}
	\item How does the choice of gain for an individual controller affect the accuracy and speed of the flexibility provision at the top layer of the cascade?
	\item How does a changing flexibility request affect convergence to a steady state for subordinate OFO controllers? 
\end{enumerate}
We expect that the time needed for flexibility provision can be decreased by increasing the gain of individual controllers. A suboptimal choice that prevents convergence to a steady-state for one controller may prevent other controllers from converging as well.   

The remainder of the paper is structured as follows: In Section~\ref{ofointro} an introduction to OFO is given with focus on the general optimization problem and the feedback nature of the approach. The hierarchical structure that was implemented is presented in Section~\ref{structure}. Section~\ref{results} explores exemplary results of the simulative study.

\section{Online Feedback Optimization}
\label{ofointro}

OFO is a control scheme that incorporates an optimization problem and its solution in a closed loop with online measurements from a physical system (the electrical grid) forming a feedback controller, as shown in Fig.~\ref{fig:ofo_complete}. In this paper, OFO is used to control the active power flow at the PCC between two grid layers while satisfying grid constraints. A possible use case for this is fulfilling a flexibility request from the superordinate grid layer in the context of ancillary services like congestion management. In the following, the resulting optimization problem that is solved by the OFO controller during one iteration is set up and the feedback structure is explained. Since the electricity grid consists of multiple voltage levels, a hierarchical coordination of the flexibility request is necessary. The resulting cascaded structure is explained in Section~\ref{structure}.

\subsection{Optimization Problem}
By integrating (\ref{eq:qp}) into the OFO controller, a constrained optimization problem (\ref{eq:genericOP}) is solved. Here, $\Phi$ denotes a cost function that can be freely chosen to represent e.g. a techno-economic objective. The result of the optimization problem is the set point vector $u$, which is also the input to the physical system. The output vector $y$ consists of the measurements from this system and the two vectors are linked by the nonlinear steady-state input to output map $y=h(u)$.
\begin{equation}
	\label{eq:genericOP}
	\begin{aligned}
		\min_{u} \enspace & \Phi(u,y)  \\
		\text{s.t.} \enspace & u \in U \\
		& y \in Y \\
		& y = h(u) \\
		\text{with} \enspace & U = \{u \in \mathbb{R}^p | Au \leq b\} \\
		\text{and} \enspace &  Y = \{y \in \mathbb{R}^n | Cy \leq d\} \\
		\text{where} \enspace & A \in \mathbb{R}^{q \times p}, \enspace b \in \mathbb{R}^{q}, \enspace C \in \mathbb{R}^{l \times n} \enspace and \enspace d \in \mathbb{R}^{l}
	\end{aligned}
\end{equation}
As long as the cost function is continuously differentiable on $\mathbb{R}$, a projected gradient descent scheme can be used to solve the presented optimization problem. The gradient of the function is updated iteratively, leading to a local minimum by going a step in the direction of the total derivative of the function during every iteration. With $h$ being continuously differentiable, the total derivative can be calculated as follows:
\begin{equation}
	\label{eq:totalderivative}
	\begin{aligned}
		\frac{d\Phi(u,y)}{du}  = & \nabla_u \Phi(u,y) \big|_{y=h(u)} \\ &  + \nabla h(u)^T \nabla_y \Phi(u,y) \big|_{y=h(u)}  \\
		 = & H(u)^T \nabla \Phi(u,y) \big|_{y=h(u)}
	\end{aligned}
\end{equation}

The sensitivity matrix $\nabla h(u)$ is the only explicit information about the system model needed by the OFO controller. Containing the steady-state sensitivities it gives information about the influence that a change in the set point vector $u$ has on the output vector $y$, as would the derivative of the power flow equations. Calculating the accurate sensitivities would require full knowledge of the grid model and state. Using constant approximate sensitivities instead yields good results, as has been experimentally proven in \cite{ortmann2020}. An entry in the sensitivity matrix is hereby calculated a-priori as follows:
\begin{equation}
	\nabla h_{i,j} = \frac{\partial h(u)_{i}}{\partial u_{j}}
\end{equation}

The calculated gradient needs to be projected onto the set of feasible points, which can be achieved by solving an internal quadratic program (QP), as defined in (\ref{eq:qp}). The result of the optimization $\hat{\sigma}_{\alpha}$ denotes the best step that can be taken into the direction of the total derivative while complying with the constraints.
\begin{equation}
	\label{eq:qp}
	\begin{aligned}
		\hat{\sigma}_{\alpha}(u,y) & = \arg \min_{w \in \mathbb{R}^p} ||w + H(u)^T \nabla \Phi(u,y) ||^2 \\
		\text{s.t.} \enspace &A(u+\alpha w) \leq b \\
		& C (y + \alpha \nabla h (u)w) \leq d \\
		\text{with} \enspace & H(u)^T = [ \mathbb{I}_p \nabla h(u)^T] \\
	\end{aligned}
\end{equation}

\subsection{Feedback Control}
\begin{figure}[tb]
	\centerline{\includegraphics{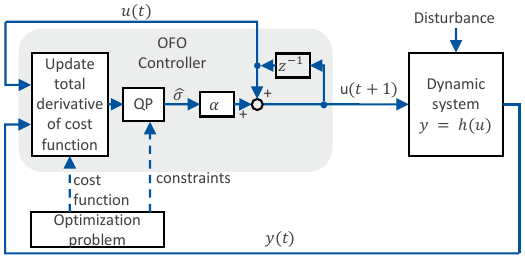}}
	\caption{Block diagram of a single OFO controller}
	\label{fig:ofo_complete}
\end{figure}

The presented optimization problem is then integrated into a feedback controller as shown in Fig.~\ref{fig:ofo_complete}.
Iteratively, the optimal inputs for the system are determined by following certain steps during one iteration of the OFO controller: 
\begin{enumerate}
	\item 	Update measurements $y$ from physical system
	\item   Calculate new $\nabla \Phi(u,y)$
	\item 	Solve QP to obtain the optimal next step
	\item 	Calculate the new set point vector $u_{t+1}$
	\item   Send new system inputs to the physical system
\end{enumerate} 	
Similar to the differential equation of a PI controller, the final input $u_{t+1}$ into the dynamic system is calculated by adding the step $\hat{\sigma}_{\alpha}$ scaled with the controller gain $\alpha$ to the set point vector of the last iteration:
\begin{equation}
	\label{eq:nextinput}
	u_{t+1} = u + \alpha \hat{\sigma}_{\alpha}
\end{equation}
Performing this calculation iteratively, asymptotic convergence to the optimal set point vector is achieved while enforcing the constraints on the system outputs. 
The feedback nature of the controller holds several advantages compared to approaches that are based on feed-forward optimization. As the sensitivity matrix is the only explicit model information needed, a detailed model of the controlled grid is not necessary. Hence, OFO is a practical solution for all cases, where system models lack accuracy or are not available. Furthermore, OFO displays robustness against model mismatch and ensures constraint satisfaction even with erroneous information. Due to the consideration of real-time measurements, OFO is able to take the current grid operating point into account. The low computational complexity allows quick satisfaction of grid constraints when used as a real-time controller  \cite{ortmann2023}.

\section{Hierarchical Flexibility Coordination}
\label{structure}
To fulfill the request for a change in active power flow at the PCC with a superordinate grid layer, a multitude of DER need to be coordinated vertically across various voltage levels. A hierarchical structure of OFO controllers inspired by the structure of the electrical grid and its system boundaries enables the provision of the requested flexibility from distribution to transmission system. 
\subsection{Cascaded OFO Controllers}
For the hierarchy of the control architecture, a directed acyclic graph $G=(V,E)$ is defined, where each of the vertices $V$ represents an individual OFO controller and the directed edges $E$ represent the communication paths between the controllers. The communication is assumed to be unidirectional during runtime, as each OFO controller can request flexibility from any subordinate OFO controller, but does not communicate with controllers on the same or on any upper level. To avoid contradicting set points, the number of incoming edges for each node is limited to one. 

\begin{figure}[tb]
	\centerline{\includegraphics{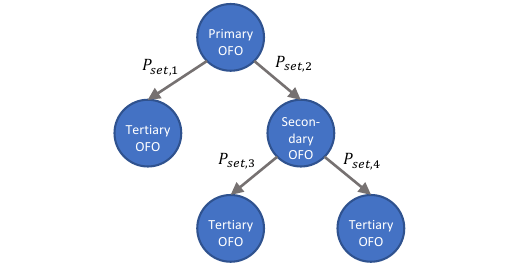}}
	\caption{Cascaded structure of individual OFO controllers as directed graph}
	\label{fig:hierarchy}
\end{figure}

Within the cascade, three different types of OFO controllers can be distinguished depending on their location in the control hierarchy. 
\begin{enumerate}
	\item Primary Controller: Only has subordinate OFO controllers, supervises the top-level grid layer
	\item Secondary Controller: Has sub- and superordinate OFO controllers, provides flexibility to superordinate grid layer
	\item Tertiary Controller: Only has superordinate OFO controllers, provides flexibility to superordinate grid layer
\end{enumerate}
The resulting cascaded structure is shown in Fig.~\ref{fig:ofo_simple} and Fig.~\ref{fig:hierarchy}. Each OFO controller is assigned to a set of controllable assets in the grid and a set of observable buses and lines, which do not overlap with the sets of other OFO controllers. One grid layer can be controlled by multiple OFO controllers as long as each asset does only receive set points from one single controller. It is not necessary that all DER and buses are controlled or observed by OFO. The robustness even in case of limited observability and other control algorithms acting on the same grid layer has been experimentally shown in \cite{klein-helmkamp2024}. 

\subsection{Optimization Problem Formulation}
The aim of providing flexibility in form of a change in active power at the PCC is formulated as an optimization problem for the OFO controllers. Due to their different tasks, the optimization problems of the controllers differ depending on their controller type as introduced before. 
The task of the primary controller is set to maintain safe grid operation regarding voltage and current limits while minimizing the difference to the planned dispatch of DER. With $F$ being the set of devices that can be actuated by an OFO controller (including DER and subordinate OFO controllers), $B$ and $N$ being the sets of observed lines and buses, the optimization problem is chosen as in (\ref{eq:opti_prob_primary}).

An explicit formulation of the power flow equations is not necessary, as these are solved implicitly by the physical system and integrated via the online measurements. This simplifies the optimization problem in contrast to other approaches. Since each controller is only responsible for part of the grid, a decoupling of the individual problems is achieved, further reducing complexity. Values for currents and voltages are obtained from measurements at the observed lines and buses.
\begin{equation} \label{eq:opti_prob_primary}
	\begin{aligned}
		\min_{P_{set},Q_{set}}   \enspace    \Phi = & \sum_{i \in F} ||P_{i,set} - P_{i,planned}||^{2} 
		\\
		\text{s. t.}  \enspace  & V_{min,n}   \leq V_{meas,n}  \leq V_{max,n} &\enspace \forall n \in N\\
		&  I_{min,i}\leq I_{meas,i} \leq I_{max,i} &\enspace \forall i \in B\\
		&  P_{min,j}   \leq P_{set,j}  \leq P_{max,j} &\enspace \forall j \in F\\
		& Q_{min,j}  \leq Q_{set,j} \leq Q_{max,j} &\enspace \forall j \in F 
	\end{aligned}
\end{equation}

Secondary and tertiary OFO controllers aim at fulfilling the flexibility request from superordinate controllers as close as possible without violating constraints within their observability area. The cost function therefore minimizes the offset from the set point $P_{set}$ as requested from the superordinate controller to the measured active power flow $P_{PCC}$ at the PCC. The constraints are equal to the ones of primary OFO controllers.
\begin{equation} \label{eq:opti_prob_secondary}
	\begin{aligned}
		\min_{P_{set},Q_{set}} \enspace \Phi = & ||P_{set,PCC} - P_{meas,PCC}||^{2} 
		\\
		\text{s. t.}  \enspace  & V_{min,n}   \leq V_{meas,n}  \leq V_{max,n} &\enspace \forall n \in N\\
		& I_{min,i}  \leq I_{meas,i} \leq I_{max,i} &\enspace \forall i \in B\\
		&  P_{min,j}   \leq P_{set,j} \leq P_{max,j} &\enspace \forall j \in F\\
		&  Q_{min,j} \leq Q_{set,j}  \leq Q_{max,j} &\enspace \forall j \in F 
	\end{aligned}
\end{equation}
\newline
\section{Results}
\label{results}
In this section the effect of the parametrization of the individual controllers in the cascade on the performance of the flexibility provision is studied. For this, a simulative case study is carried out on a three-level test system. The test system consists of a high voltage, a medium voltage and a low voltage layer with DER and loads connected according to table \ref{tab:test_system}. Each layer is controlled by one OFO controller, of secondary or tertiary type. The primary controller is assumed to request a fixed operating point of $P_{PCC}=-120\,MW$ at the PCC between EHV and HV layer which shall be provided using the controllable DER and loads connected to the distribution system. For the simulative study a stationary load flow calculation is executed, assuming a constant operating point of the grid, which is only changed through the OFO controllers. External disturbances are neglected here, as robustness of the approach has been shown in other work \cite{klein-helmkamp2023}. 
\begin{table}[h]
	\begin{center}
		\caption{Controller Cascade and Test System}
		\begin{tabular}{||l c c c c||} 
			\hline
			Controller & \parbox{1cm}{\centering Type} & \parbox{1cm}{\centering Voltage Level} & \parbox{1.5cm}{\centering Controllable DER [MW] }& \parbox{1.5cm}{\centering Controllable Load [MW]}\\ 
			[1ex] 
			\hline\hline
			$OFO_{1}$ & Secondary & HV & 45 & 190 \\ 
			\hline
			$OFO_{2}$ & Secondary & MV & 6.75 & 8.4 \\ 
			\hline
			$OFO_{3}$ & Tertiary & LV & 0.062 & 0.21 \\
			\hline
		\end{tabular}
		\label{tab:test_system}
	\end{center}
\end{table}

\subsection{Uniform Choice of Controller Gain $\alpha$}
\begin{figure}[tb]
	\centerline{\includegraphics{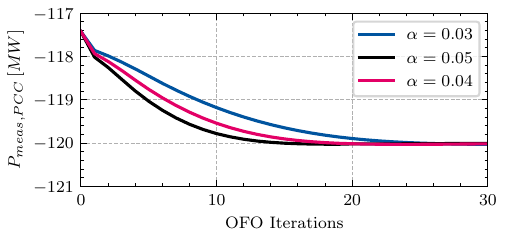}}
	\caption{Active power flow at the EHV/HV PCC depending on the controller gain of the OFO controllers}
	\label{fig:fixedalpha}
\end{figure}

Due to its iterative nature, the measured active power flow at the PCC controlled by an OFO controller converges stepwise to the requested set point. The convergence speed is thereby influenced by the controller gain. This is because every step, which is added to the set point vector $u$ after an OFO iteration, is scaled with $\alpha$ (compare (\ref{eq:nextinput})). Figure~\ref{fig:fixedalpha} shows the power flow at the PCC between the HV layer controlled by $OFO_1$ and the external EHV layer. In a first study the same $\alpha$ was chosen for every controller in the grid to showcase the dependency between convergence speed and tuning. As can be seen, with a higher gain, the requested set point of $P_{set} = -120\,MW$ is reached faster. For $\alpha = 0.05$ the measured power flow at the PCC reaches the requested set point after 15 iterations, while for $\alpha = 0.03$ the set point is not reached even after 24 iterations. Hence, a careful choice of the controller gain has a significant impact on the time needed for the flexibility provision at the top layer of the cascade. For even higher values, unwanted oscillations can occur, as shown in Section~\ref{sec:instability}.   

\subsection{Set Point Tracking}
\begin{figure}[tb]
	\centerline{\includegraphics{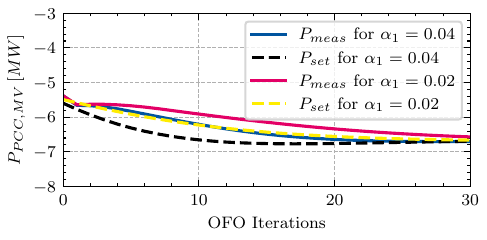}}
	\caption{Active power flow to the requested set point from superordinate OFO controller}
	\label{fig:setpointtracking}
\end{figure}

To evaluate the influence of a single controller on the cascade only the gain of $OFO_1$ is varied, while the parameters of the controllers $OFO_2$ and $OFO_3$ are not changed. Fig.~\ref{fig:setpointtracking} shows the power flows at the PCC between HV and MV layer for $\alpha_1=0.04$ and $\alpha_1 = 0.02$. The gains of the other two controllers are kept constant at $\alpha_2 = \alpha_3 = 0.025$. The dashed lines represent the set point requested by $OFO_1$ from $OFO_2$. As can be seen, the requested set point changes at every iteration until reaching a steady state. The measured power flow, which is controlled by $OFO_2$, converges to the requested set point until it also reaches a steady state. Hence, for the considered case, the parameter choice of the superordinate OFO controller also influences the provision of flexibility through the subordinate one.

This showcases the iterative nature of the approach. In every iteration the superordinate controller calculates a new set point for all its controlled DER and subordinate OFO controllers. A smaller gain signifies slower convergence to a steady state regarding the calculated set points. The optimization problem of the subordinate controller is updated at every iteration with the new set point. The action of every secondary or tertiary controller is therefore directly dependent on the convergence behavior of its superordinate controller and therefore of the respective gain. 

\subsection{Influence of Suboptimal Parameterization}
\label{sec:instability}
\begin{figure}[h]
	\centerline{\includegraphics{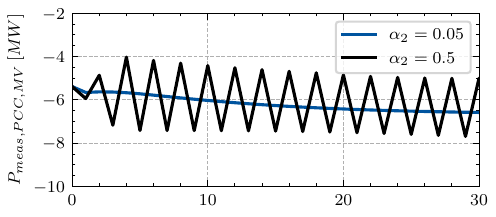}}
	\centerline{\includegraphics{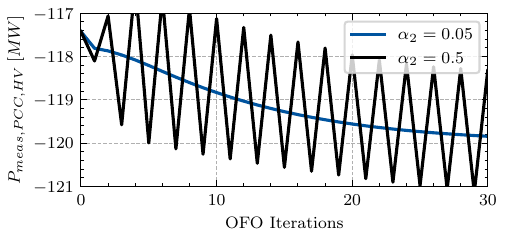}}
	\caption{Active power flow of PCC between EHV/HV layer ($P_{meas,PCC,HV}$) and HV/MV layer ($P_{meas,PCC,MV}$) depending on choice of $\alpha$ for OFO\textsubscript{2}}
	\label{fig:instability}
\end{figure}

In Fig.~\ref{fig:instability} the measured active power flow at the PCC between HV and MV layer and at the PCC between EHV and HV layer is displayed. For $OFO_1$ and $OFO_3$ the parameters are kept constant at $\alpha_1=\alpha_3=0.02$ while the gain of $OFO_2$, which controls the MV layer, is varied between $\alpha_2=0.05$ (blue graph) and $\alpha_2=0.5$ (black graph). For the latter, an oscillation of the measured power flow as an example of unwanted behavior can be observed for both PCCs, thus the OFO controller is not able to achieve convergence to a steady state for the observed time frame. This behavior can also be observed for single OFO controllers that are not implemented as part of a cascade if the gain $\alpha$ is chosen too high. In this case the gain of controller $OFO_2$ is chosen in such a way to trigger these oscillations. As can be seen from Fig.~\ref{fig:instability}, not only the measured power flow at the PCC controlled by $OFO_2$ starts oscillating, but also the superordinate $OFO_1$ displays oscillating behavior although its parameterization is the same for both cases (black and blue graph). An attentive tuning of individual controllers therefore has significant impact on the accuracy and speed of flexibility provision and should be done considering the cascaded structure.  

\section{Conclusion}
In this paper the influence of the tuning of individual controllers in a cascaded structure based on online feedback optimization was studied. The implemented structure consists of hierarchically connected individual OFO controllers, each controlling their own grid layer. The algorithm is based on an optimization problem integrated into a feedback controller. Closing the loop with the dynamical system provides robustness against model mismatch and allows fast provision of flexibility while ensuring grid constraints. The optimization problem is set up to actuate DER connected to the respective grid layer with the aim of changing the active power flow at the PCC to the superordinate grid layer. The implemented cascade is studied on a three-level distribution system consisting of low to high voltage levels. It shows high suitability for vertical flexibility coordination. 

To showcase interactions between the individual controllers within the cascade, the gain of individual controllers was varied and the effects on the flexibility provisions observed. The results show a high interdependency between the parameterization of individual controllers and the behavior of other controllers in the hierarchy. Tuning can either improve overall convergence speed or lead to oscillations in the worst case. Oscillations caused by wrong tuning of one controller can also appear on other grid layers, preventing flexibility provision entirely. A careful tuning of individual controllers is therefore essential for a reliable function of the entire cascade and should be done considering the achievement of a stable convergence behavior on one hand and a high convergence speed on the other hand. Apart from testing different scenarios in general, further aspects can be taken into account in future research. These can include timescale separation, to further optimize controller interaction, as well as effects of erroneous information resulting from faulty communication between the controllers. Furthermore, a comparative analysis should be conducted to benchmark the performance of the cascaded OFO structure against other approaches.

\end{document}